\documentclass[aps,prl,twocolumn,10pt,superscriptaddress,showpacs,showkeys]{revtex4-1}
\usepackage[T1]{fontenc}
\usepackage{lmodern}
\usepackage{graphicx}

\begin{document}


\title{Valence Instability of YbCu$_2$Si$_2$ through its quantum critical point}


\author{A. Fernandez-Pa\~nella}
\affiliation{SPSMS, UMR-E CEA / UJF-Grenoble 1, INAC, 38054 Grenoble, France}
\author{V. Bal\'{e}dent}  
\affiliation{Synchrotron SOLEIL, L'Orme des Merisiers, Saint-Aubin, BP~48, 91192 Gif-sur-Yvette Cedex, France}  
\author{D. Braithwaite}
\email [Corresponding author: ] {daniel.braithwaite@cea.fr}
\affiliation{SPSMS, UMR-E CEA / UJF-Grenoble 1, INAC, 38054 Grenoble, France}
\author{L. Paolasini}
\affiliation{ESRF, 6 Rue Jules Horowitz, BP 220, 38043 Grenoble Cedex, France}
\author{R. Verbeni}
\affiliation{ESRF, 6 Rue Jules Horowitz, BP 220, 38043 Grenoble Cedex, France}
\author{G. Lapertot}
 \affiliation{SPSMS, UMR-E CEA / UJF-Grenoble 1, INAC, 38054 Grenoble, France}
\author{J.-P.~Rueff}  
\affiliation{Synchrotron SOLEIL, L'Orme des Merisiers, Saint-Aubin, BP~48, 91192 Gif-sur-Yvette Cedex, France}  
\affiliation{Laboratoire de Chimie Physique--Mati\`ere et Rayonnement, CNRS-UMR~7614, UPMC, 75005, Paris, France}



\begin{abstract}

We report Resonant inelastic x-ray scattering measurements (RIXS) in YbCu$_2$Si$_2$ at the Yb L$_{3}$ edge under high pressure (up to 22 GPa) and at low temperatures (down to 7 K) with emphasis on the vicinity of the transition to a magnetic ordered state. 
We find a continuous valence change towards the trivalent state with increasing pressure but with a pronounced change of slope close to the critical pressure. Even at 22 GPa the Yb$^{+3}$ state is not fully achieved. The pressure where this feature is observed  decreases as the temperature is reduced to 9 GPa  at 7K, a value close to the critical pressure (\itshape{p\normalfont{$_c$}}\normalfont $\approx$ 7.5 GPa) where magnetic order occurs. The decrease in the valence with decreasing temperature previously reported at ambient pressure is confirmed and is found to be enhanced at higher pressures.  We also compare the f electron occupancy between YbCu$_2$Si$_2$ and its Ce-counterpart, CeCu$_2$Si$_2$.
 
\end{abstract}


\maketitle{}

One of the major issues not fully resolved in the study of rare-earth (RE) intermediate-valence (IV) compounds is the interplay between magnetic and valence instability, especially when the system is driven by the application of an external parameter as pressure towards a magnetic quantum critical point (QCP), where strong spin and/or valence fluctuations are expected to arise. 
Especially noteworthy are the IV REM$_2$X$_2$ (where RE=Ce,Yb, M=transition metal, X=Si,Ge) compounds widely studied during the past decades owing to the interesting physical properties they exhibit as heavy fermion behavior, different types of magnetic order, superconductivity and non-Fermi liquid behavior. All these phenomena are based on the non-integer occupancy of the 4\itshape f \normalfont orbital and to the competition among the significant energy scales of these systems, i.e crystal electric field effect (CEF), Kondo effect and Ruderman-Kittel-Kasuya-Yoshida (RKKY) interaction.
One important aspect to get a deeper understanding of these phenomena, is a systematic comparison between cerium and ytterbium systems. Yb is often considered to be the hole equivalent of cerium. Pressure tends to drive Yb from its nonmagnetic Yb$^{2+}\!(4f^{14})$ state to a magnetic Yb$^{3+}\!( 4f^{13})$ state. As pointed out by Flouquet and Harima \cite{harima09} there are however significant differences between the 2 families. In Yb the deeper localisation of the $4f$ electrons, and the larger spin orbit coupling lead to a different hierarchy of the significant energy scales. One consequence is that whereas in Ce the valence change will be quite restricted, in Yb applying pressure is expected to induce larger changes of the valence.
YbCu$_2$Si$_2$ is the ideal prototype Yb system, apparently behaving as expected. The application of pressure produces a mirror image of a typical cerium phase diagram, with decreasing Kondo temperature, increasing magnetic fluctuation contributions to the resistivity, and for pressures above 8 GPa magnetic order. Additional interest is the recent discovery of ferromagnetism.\cite{fernandez11} Furthermore, a recent study has shown that for its Ce counterpart CeCu$_2$Si$_2$\cite{rueff11}, a significant valence crossover is induced with pressure, which might support recent theoretical developments that valence fluctuations considerably enhance superconductivity. It is therefore of particular interest to follow how the Yb valence changes as pressure allows to scan the full phase diagram from deep in the paramagnetic state through the critical pressure and into magnetic order. As the energy scales and the magnetic ordering temperature are low, it is important to combine the extreme conditions of high pressure and low temperature. 
The recent development of truly bulk sensitive and resonant spectroscopic techniques has significantly improved the accuracy of the study of the electronic structure under high pressure conditions \cite{Rueff2010,rueff06,moreschini07}. The measurement of the RE valence and how it evolves under high pressure and at low temperatures turns out to be a relevant tool because it quantifies the hybridization between the \itshape f \normalfont electron with the conduction band \cite{dallera02,moreschini07,kummer11}. However, up to date, few experimental results have been reported so far under such extreme conditions \cite{rueff11}. 
Experiments performed at ambient pressure and low temperatures or at room temperature and high pressures are more accessible. Some results have been reported using different spectroscopic techniques as x-ray absorption (XAS), partial-fluorescence yield mode (PFY) and resonant x-ray emission (RXES) for different REM$_2$X$_2$ systems as YbCu$_2$Si$_2$ \cite{lawrence94,moreschini07}, YbNi$_2$Ge$_2$ and YbPd$_2$Si$_2$\cite{yamaoka10}. At ambient pressure, all of them are in a mixed valent state with a valence v = 2.88, 2.91 and 2.94 respectively. The valence-temperature dependence down to 20 K shows a continuos decrease for the three compounds of $\Delta$v = 0.06, 0.1 and 0.06 respectively. When pressure is applied (at 300 K) the Yb ion is driven towards its trivalent configuration, even though it does not fully achieved it, by an amount of $\Delta$v = 0.07 for YbNi$_2$Ge$_2$ and $\Delta$v = 0.05 for YbPd$_2$Si$_2$. This trend was already proposed by theory based on thermodynamic arguments which pointed out a higher stability of the trivalent state at sufficient high pressures.\cite{borje75} All these measurements were taken too far from a critical region, either in pressure or temperature, in order to give accurate details of the strong correlations among the low energy interactions in these compounds. 
In this paper we report a detailed study of the Yb valence from RIXS measurements in the moderate heavy fermion YbCu$_2$Si$_2$ ($\gamma \approx$135 mJ mol$^{-1}K^{-2}$)\cite{Sales76} at several points of the \itshape{p-T} \normalfont phase diagram (p$_{max}$= 22 GPa, T$_{min}$= 7 K) and for the first time, very close to the ferromagnetic order state\cite{fernandez11}, as summarized in Fig.1. Red\cite{Colombier2009} and black stars\cite{Yadri98} correspond to the temperature at the maximum of the resistivity curve (T$_{Max}$) which is considered to be about the same order of magnitude of the Kondo temperature, T$_K$, which strongly decreases with pressure up to 15 GPa. 
\begin{figure}[htbp]
\includegraphics [width=6cm]{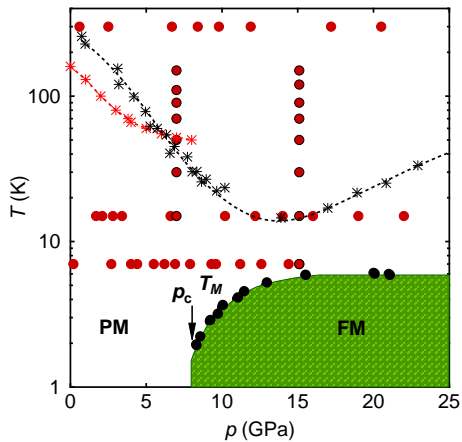}
\caption{\footnotesize(Color online) \itshape{p-T} \normalfont phase diagram of YbCu$_2$Si$_2$ (adapted from Ref.~\cite{fernandez11} and for $p$>13 GPa from Ref. \cite{Yadri98}). T$_M$ and p$_c$ are the ferromagnetic order temperature and the critical pressure respectively. Filled red points correspond to the positions where the valence was measured. Points with a black stroke correspond to the valence-temperature depen\-dence at 7 and 15 GPa. The valence-pressure dependence was measured at 300, 15 and 7 K. T$_{Max}$(red \cite{Colombier2009} and black stars \cite{Yadri98}) is related to the Kondo temperature (see details in the text).}
\end{figure}
We have used high quality single crystals grown by in-flux method (using MgO crucibles) as described in detail elsewhere\cite{Colombier2009}. Measurements were performed at the beamline ID16 of the European Synchrotron Radiation Facility (ESRF, Grenoble). The undulator beam was monochromatized with a Si(111) monochromator and focused to a size of 40 $\mu$m (Vertical) and 100 $\mu$m (Horizontal) at the sample position. The scattered x-rays were analyzed by a Rowland circle spectrometer equipped with a spherically bent Si (620) crystal. The energy resolution was about 1.5 eV. A sample $\sim$20 $\mu$m thick was loaded in a membrane-driven diamond anvil cell (DAC) with silicon oil as a pressure transmitting medium. A helium circulation cryostat was used to measure at low temperatures down to 7 K. Pressure in the DAC was estimated by the fluorescence of the ruby chip placed in the pressure chamber.
\\The main advantage of working in resonant regime (RIXS) is the possibility to selectively enhance the Yb$^{+2}$ $(2p^6f^{14}5d^0)$ or Yb$^{+3}$ $(2p^6f^{13}5d^1)$ component by an appropriate choice of the incident photon energy $h\nu_{in}$ compare to PFY-XAS experiments which is also a core-hole and "high resolution" spectroscopy technique where a specific fluorescence channel is measured. Smaller relative changes  between the intensities of the two components can be better distinguished in the RIXS spectra and therefore a higher accuracy in the estimation of the valence is obtained.\cite{moreschini07} 
\\The RIXS spectra of YbCu$_2$Si$_2$ are summarized in Fig.2. The two main features at an energy transfer, E$_t=h\nu_{in} - h\nu_{out}$, of 1525.5 and 1530.5 eV correspond to transitions from the Yb$^{+2}$ and Yb$^{+3}$ components of the initial mixed valent ground state $(\left|g\right\rangle = a\left|f^{13}\right\rangle + b\left|f^{14}\right\rangle)$. They are separated by $\approx5$ eV which corresponds to the Coulomb repulsion between the $4f$ hole and the $3d$ core hole in the final excited states $(3d^9f^{14}\epsilon d)$ and $(3d^9f^{13})\epsilon d$ ($\epsilon d$ is an electron added to a valence band of d character). We measured at a $h\nu_{in} = 8.9404$ keV to enhance the Yb$^{+2}$ component. This value was chosen after the PFY-XAS spectra were acquired to monitored the Yb $L\alpha_1 (2p\rightarrow3d)$ line while $h\nu_{in}$ was swept across the Yb $L_3$ ${(2p\rightarrow5d)}$ absorption edge and we could see for which $h\nu_{in}$ the intensity for the Yb$^{+2}$ was maximized. For convenience, all spectra are normalized to the maximum of their respective Yb$^{+2}$ signal and plotted versus E$_t$. There is a third weak low-energy feature that shows up only at high pressures (p> 7 GPa) around E$_t = 1520$ eV. It corresponds to a quadrupole-allowed ($E2$) $2p^64f^13\rightarrow2p^54f^14$ transition of Yb$^{+3}$ better distinguished with PFY-XAS \cite{moreschini07}. 
Fig.2(a) shows the spectra measured at 7 K, the lowest temperature that could be reached due to technical limitations of the membrane-driven pressure cells and in the cooling power of the cryostat. A clear increase of the Yb$^{+3}$ intensity under pressure compare to Yb$^{+2}$ is observed while the weak $E2$ feature stays mostly unchanged. At 15 K, in Fig.2(b) and at 300 K (not shown here) a similar trend under pressure is noticed. This spectral weight transfer from Yb$^{+2}$ towards Yb$^{+3}$ is in accordance with the delocalization process of a $4f$ electron under pressure. It has been reported\cite{moreschini07} that the valence of YbCu$_2$Si$_2$ at ambient pressure decreases when the system is cooled. This tendency is also observed in our results under pressure. As can be clearly seen in Fig.2(c) and (d) at 7 GPa and 15 GPa respectively, the intensity of the Yb$^{+3}$ peak decreases with temperature, i.e, the system is driven to the divalent state.
\begin{figure}[htbp]
\includegraphics [width=7cm]{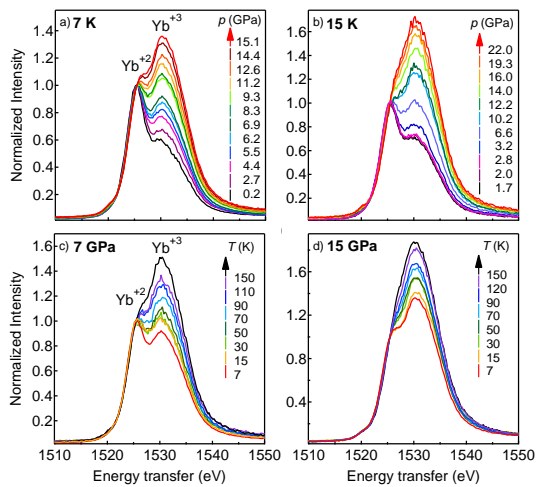}
\caption{\footnotesize(Color online) Summary of RIXS spectra at (a) 7 K, (b) 15 K at selected pressures; (c) and (d) correspond the spectra at 7 GPa and 15 GPa respectively for different temperatures.}
\end{figure}
A quantitative estimation of the Yb valence from the spectral data was calculated using the following expression $v = 2 + I^{+3}/\left(I^{+2}+I^{+3}\right)$. The integrated $I^{+2}$ and $I^{+3}$ intensities where evaluated by fitting a superposition of two Gaussian functions to our data, one for each spectral contribution, and also an arctangent function to fit the background. The weak  $E2$ transition was not included in this analysis. The values are reported in Fig.3. 
\begin{figure}[htbp]
\includegraphics [width=8.5cm]{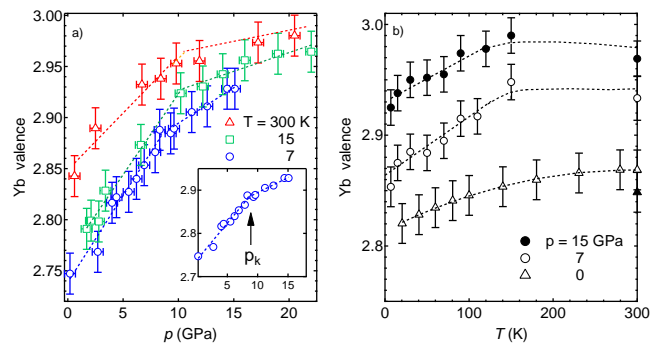}
\caption{\footnotesize(Color online) (a) valence-pressure dependence of YbCu$_2$Si$_2$ at 7, 15 and 300 K. The inset shows the values at 7 K. The arrow points the pressure where a kink is observed ($p_k$).(b) Valence-temperature dependence at ambient pressure\cite{moreschini07}, 7 and 15 GPa. The full triangle point is the value we obtained at 300 K and ambient pressure. Dashed lines are guides to the eyes.}
\end{figure}
The valence-pressure dependence in Fig.3(a) clearly shows how the trivalent state is favoured under pressure and gives a precise picture of how the Yb valence changes  at temperatures and pressures close to the critical region. The valence increases monotonously with pressure for all temperatures we measured: 300 K, 15 K and 7 K with a $\Delta$v (up to 15 GPa) of 0.12, 0.15 and 0.18 respectively. The main result of this work is a distinctive change of slope or kink which occurs at a pressure $p_k$ above which a significant decrease of the rate $\partial v/\partial p$ for p > $p_k$ is found. An estimation of $p_k$ (as indicated with an arrow in the inset of Fig.3(a) for the values at 7 K) reveals that it decreases from $p_k\approx$ 11 GPa (300 K) to $\approx$ 9 GPa (7 K) approaching $p_c$ . Above $p_k$ the valence keeps increasing slowly under pressure and it does not achieve the fully trivalent state even at 22 GPa, the highest pressure achieved. Because the lowest temperature possible was about 7 K, we could not measure the valence in the magnetically ordered state, however for p > 12 GPa, the system was very close to this (see Fig.1), only about 1 K above T$_M$. Although we cannot exclude that the valence will jump to 3 at the onset of magnetic order, this seems unlikely. An apparent value of less than 3 could also arise from inhomogeneity: there is evidence that the onset of magnetic order occurs as a first order transition, and a previous M\"{o}ssbauer study found a magnetic and a non-magnetic component just above $p_c$. However this would imply that this phase separation exists even at 22 GPa, which also seems extremely unlikely. We therefore conclude that in this high pressure region of the phase diagram magnetism sets in for a valence value less than 3, so in the mixed valency state.

 Fig.3(b) shows the valence-temperature dependence at three different pressures: ambient pressure\cite{moreschini07}, at 7 $\pm$ 1 GPa and at 15 $\pm$ 1 GPa, very close and well above p$_c$ respectively. The values at 300 K are estimated from Fig.3(a). The valence-temperature behavior is, at first sight, roughly the same for all pressures. At ambient pressure, slightly valence changes down to 200 K are detected but below this temperature the slope $\left|\partial v/\partial T\right|$ is considerably enhanced. The decrease of the valence is much more pronounced at high pressure, but the temperature below which this decrease occurs does not seem to change under pressure. No significant differences can be discerned between 7 GPa and 15 GPa. This implies that this decrease is not related directly to the formation of the heavy fermion state, as although at ambient pressure the Kondo temperature is estimated to be of the same order as the temperature where this feature is found, at higher pressures it decreases strongly (see Fig.1).
\begin{figure}[htbp]
\includegraphics [width=6cm]{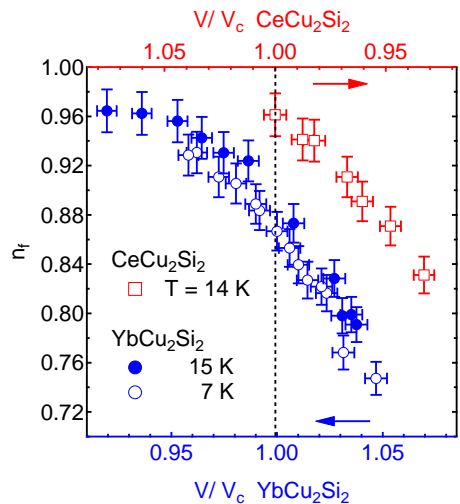}
\caption{\footnotesize (Color online) f electron and hole occupancy of CeCu$_2$Si$_2$ (upper axis) and YbCu$_2$Si$_2$ (low axis) respectively versus their molar volume normalized to their critical molar volume, V$_c$ (see details in text). Blue and red arrows indicate the  direction of increasing applied external pressure.}
\end{figure}
Recent valence measurements carried out in CeCu$_2$Si$_2$ in similar extreme conditions\cite{rueff11} offers the valuable and rare opportunity to compare the 4f occupancy (n$_f$) under pressure and at low temperatures between a Ce-based compound and its Yb-counterpart. CeCu$_2$Si$_2$ is very sensitive to small differences in stoichiometry which can lead at ambient pressure to find samples with an antiferromagnetic (AF) transition around 0.7 K or a superconductivity below 0.65 K or the combination of the two\cite{steglich79,bellarbi,yuan03,fujiwara} . However it is generally accepted that at ambient pressure CeCu$_2$Si$_2$ is very close to an AF QCP. Under pressure superconductivity is enhanced and shows a two-dome shape (the first dome is located at p$_c$ and the second one at p$_v\approx$ 4.5 GPa). The strong magnetic fluctuations that arises near p$_c$ are held responsible for the electron pairing near the magnetic-QCP. For the second dome at p$_v$ the f-electron occupancy and associated valence fluctuations might play an important role in superconductivity\cite{rueff11,alexander07}.
As YbCu$_2$Si$_2$ (bulk modulus, $B_0$=168 GPa)\cite{sanchez00} is harder than CeCu$_2$Si$_2$ ($B_0$=112 GPa)\cite{tsuduki05} it is more appropriate to compare their 4f electron (Ce) and hole (Yb) occupancy, n$_f$, respect their molar volume change by using the p-V relation $V=\!V_0 \exp\left(-\kappa\Delta p\right)$ rather than the applied external pressure. The molar volume at ambient pressure and at low temperatures, $V_0$, was calculated using the $a$ and $c$ values from ref.\cite{neumann}. The molar volume change is normalized to the critical molar volume for each compound $V_c=146.012$ \AA$^3$ and $165.324$ \AA$^3$ for YbCu$_2$Si$_2$ and CeCu$_2$Si$_2$ respectively. The upper axis has been shifted in order to align the critical molar volume for both compounds as indicated by the vertical dotted line. The blue and red arrows indicate the increasing sense for pressure. For a $\Delta p \approx$ 8 GPa the variation of the molar volume is about 6.7\% in CeCu$_2$Si$_2$ while 4.5\% for YbCu$_2$Si$_2$ and their $\Delta n_f$ change is about 13.5\% and 15.4\% respectively. This result bears out what it was suggested theoretically\cite{harima09}: the differences between Yb and Ce-based systems which leads to a different hierarchy in the main energy scales (T$_K$, CEF) might allow a wider scan in the valence between the divalent and trivalent states in the former compounds. Further studies in other compounds would be of great interest in order to clarify if this is a general tendency in Ce and Yb systems or a particular trend of RECu$_2$Si$_2$ family (with RE=Ce, Yb). The effect would be even more dramatic if we compare $\Delta n_f$ for a same $\Delta V\!/\!V_c$. From these results it is tempting to conclude that in YbCu$_2$Si$_2$ valence fluctuations will play a significant role, and that superconductivity should be sought not at the critical pressure but somewhat below this. This is however an extreme simplification, and other differences should be taken into account, not least the ferromagnetic nature of the magnetic order, and possible first order type of the critical point in YbCu$_2$Si$_2$.
\\In conclusion, we have studied by resonant x-ray spectroscopy at the Yb $L_3$ edge the valence properties of YbCu$_2$Si$_2$ under high pressures and low temperatures, near the critical region and the magnetic order phase. A significant and continuous change under pressure of the f electron occupancy is observed, with a distinctive change of slope close to the critical pressure where magnetic order occurs. However, the fully trivalent state is not yet achieved at the higher pressure of this study, which implies that YbCu$_2$Si$_2$ remains in a mixed-valency state for a wide range of its phase diagram, also in the magnetic ordered phase. The results also show that a larger valence change under pressure is achieved in the Yb-based system compound compare to its Ce counterpart but further studies are needed to verify if it is a general trend between Ce and Yb systems.


\begin{acknowledgments}
We thank J. Flouquet, H. Harima and S.Burdin for fruitful discussions. This work was supported by the French ANR agency within the project Blanc PRINCESS

\end{acknowledgments}

\bibliography{bibliorixs}

\end{document}